\documentclass{ws-rv9x6}  

\usepackage{latexsym}
\usepackage{graphicx}

\newcommand{\cc}{cosmological constant}

\def\rb{\right)}
\def\lb{\left(}  
\def\eq#1{{Eq.~(\ref{#1})}}

\def\frab#1#2{\left(\frac{#1}{#2}\right)}         
 \def\fra#1#2{\frac{#1}{#2}}

\begin{document}

\setcounter{chapter}{0}

\chapter{UNDERSTANDING OUR UNIVERSE: CURRENT STATUS AND OPEN ISSUES}

\markboth{T. Padmanabhan}{Understanding our Universe: Current status and open issues}

\author{T. Padmanabhan}

\address{IUCAA
P.O. Box 4,
 Pune University Campus, \\
Ganeshkhind,
Pune - 411 007, India.\\
E-mail: nabhan@iucaa.ernet.in}

\begin{abstract}
Last couple of decades have been the golden age for cosmology. High quality data  confirmed the broad paradigm of standard
 cosmology but have thrusted upon us a preposterous
composition for the universe which defies any simple explanation, thereby posing probably the greatest challenge
theoretical physics has ever faced. Several aspects of these developments are critically reviewed, concentrating on conceptual issues and open questions.
\end{abstract}



\section{Prologue: Universe as a physical system}

Attempts to understand 
the behaviour of our universe by applying the laws of physics 
lead to difficulties which have no parallel in the application
of laws of physics to systems of  more moderate scale --- like atoms, solids or even galaxies. 
 We have only one  universe available for study, which itself is evolving in time; hence, different epochs in the past history
of the universe are unique and have occurred only once. Standard rules of science, like repeatability, statistical
stability and predictability cannot be applied  to the study of the entire universe in a naive manner. 

The obvious procedure will be to start with the current state of 
the universe and use the laws of physics to study its past and future. Progress in this attempt is limited because
  our  understanding of physical processes
  at energy scales above 100 GeV or so  lacks direct experimental support.
What is more, cosmological observations suggest that nearly 95 per cent of
the matter in the universe is of types which have not been seen in the
 laboratory; there is also  indirect, but definitive,
evidence to suggest that  nearly 70 per cent of the  matter present in the universe 
exerts negative pressure.

These difficulties --- which are unique when we attempt to apply the 
laws of physics to an evolving universe --- require  the cosmologists to proceed in a multi faceted manner. The standard
 paradigm is based on the idea that the universe was
reasonably homogeneous, isotropic  and fairly featureless --- 
except for small fluctuations in the energy density ---
at sufficiently early times. It is then possible to integrate the equations describing
   the universe 
 forward in time. The results will depend on only a small number (about half a dozen) of parameters
describing the composition of the universe, its current expansion rate
and the initial spectrum of density perturbations.  Varying these parameters allows us to construct a library 
of evolutionary models for the universe which could then be 
compared with observations in order to restrict the parameter space. 
We shall now describe some of the details in this approach.

\section{The Cosmological Paradigm}

 Observations show that the universe is fairly homogeneous and isotropic at scales
larger than about $150h^{-1}$ Mpc,  where 1 Mpc $\approx 3\times 10^{24}$ cm is a convenient unit for extragalactic astronomy and $h\approx 0.7$ characterizes\cite{h} the current rate of expansion of the universe in dimensionless form. (The mean distance between galaxies is about 1 Mpc while the size of the visible universe is about $3000 h^{-1}$ Mpc.) The conventional --- and highly successful --- approach to cosmology
separates the study of large scale ($l\gtrsim 150h^{-1}$ Mpc) dynamics of the universe from the issue of structure formation at smaller scales. The former is modeled by  a homogeneous and isotropic distribution of energy density; the latter issue is
addressed  in terms of gravitational instability which will amplify the small perturbations in the energy density, leading to the formation of structures like galaxies.

In such an approach, the expansion of the background universe is described by a single function of time $a(t)$
which is governed by the equations (with $c=1$): 
\begin{equation} 
\frac{\dot a^2+k}{a^2} =\frac{8\pi G\rho}{3};\qquad d(\rho a^3)=-pda^3
\label{frw}
\end{equation}
The first one relates expansion rate to the energy density $\rho$ and $k=0,\pm 1$ is a parameter which characterizes the spatial curvature of the universe. 
 The second equation, when coupled with the equation of state 
$p=p(\rho)$ which relates the pressure $p$ to the energy density, determines the evolution of energy density  $\rho=\rho(a)$ in terms of the expansion factor of the universe.
 In particular if $p=w\rho$ with (at least, approximately) constant $w$ then, $ \rho \propto a^{-3(1+w)}$ and (if we  further assume $k=0$, which is strongly favoured by observations) the first equation in Eq.(\ref{frw}) gives
$ a \propto t^{2/[3(1+w)]}$. We will also often use the redshift $z(t)$, defined as $(1+z)=a_0/a(t)$ where the subscript zero
denotes quantities evaluated at the present moment.
 
 It is convenient to measure
the energy densities of different components in terms of a \textit{critical energy density} ($\rho_c$) required to make $k=0$ at the present epoch. (Of course, since $k$ is a constant,
it will remain zero at all epochs if it is zero at any given moment of time.) From Eq.(\ref{frw}), it is clear that $\rho_c=3H^2_0/8\pi G$ where $H_0\equiv (\dot a/a)_0$ --- called the Hubble constant ---
is the rate of expansion of the universe at present.  The variables $\Omega_i\equiv \rho_i/\rho_c$ 
will give the fractional contribution of different components of the universe ($i$ denoting baryons, dark matter, radiation, etc.) to the  critical density. Observations then lead to the following results:

(1) Our universe has $0.98\lesssim\Omega_{tot}\lesssim1.08$. The value of $\Omega_{tot}$ can be determined from the angular anisotropy spectrum of the cosmic microwave background radiation (CMBR; see Section \ref{sec:tempcmbr})  and these observations (combined with the reasonable assumption that $h>0.5$) show\cite{cmbr,kanduhere}  that we live in a universe
with critical density, so that $k=0$.

(2) Observations of primordial deuterium produced in big bang nucleosynthesis (which took place when the universe
was about few minutes in age) as well as the CMBR observations show\cite{baryon}  that  the {\it total} amount of baryons in the
universe contributes about $\Omega_B=(0.024\pm 0.0012)h^{-2}$. Given the independent observations\cite{h} which fix $h=0.72\pm 0.07$, we conclude that   $\Omega_B\cong 0.04-0.06$. These observations take into account all baryons which exist in the universe today irrespective of whether they are luminous or not. \textit{Combined with previous item we conclude that
most of the universe is non-baryonic.}

(3) Host of observations related to large scale structure and dynamics (rotation curves of galaxies, estimate of cluster masses, gravitational lensing, galaxy surveys ..) all suggest\cite{dm} that the universe is populated by a non-luminous component of matter (dark matter; DM hereafter) made of weakly interacting massive particles which \textit{does} cluster at galactic scales. This component contributes about $\Omega_{DM}\cong 0.20-0.35$ and has the simple equation of state $p_{DM}\approx 0$. (In the relativistic theory, the pressure $p\propto m v^2$ is negligible
compared to energy density $\rho\propto mc^2$ for non relativistic particles.). The second equation in Eq.(\ref{frw}), then gives $\rho_{DM}\propto a^{-3}$  as  the universe expands which arises from the evolution of number density of particles: $\rho=nmc^2
\propto n\propto a^{-3}.$

(4) Combining the last observation with the first we conclude that there must be (at least) one more component 
to the energy density of the universe contributing about 70\% of critical density. Early analysis of several observations\cite{earlyde} indicated that this component is unclustered and has negative pressure. This is confirmed dramatically by the supernova observations (see Ref.~\refcite{sn}; for a critical look at the data, see Ref.~\refcite{tptirthsn1}).  The observations suggest that the missing component has 
$w=p/\rho\lesssim-0.78$
and contributes $\Omega_{DE}\cong 0.60-0.75$. The simplest choice for such \textit{dark energy} with negative pressure is the cosmological constant which is  a term that can be added to Einstein's equations. This term acts like  a fluid with an equation of state $p_{DE}=-\rho_{DE}$; the second equation in Eq.(\ref{frw}), then gives $\rho_{DE}=$ constant as universe expands.

(5) The universe also contains radiation contributing an energy density $\Omega_Rh^2=2.56\times 10^{-5}$ today most of which is due to
photons in the CMBR. 
The equation of state is $p_R=(1/3)\rho_R$; the second equation in Eq.(\ref{frw}), then gives $\rho_R\propto a^{-4}$. Combining it with the result
$\rho_R\propto T^4$ for thermal radiation, it follows that
 $T\propto a^{-1}$.
Radiation is dynamically irrelevant today but since $(\rho_R/\rho_{DM})\propto a^{-1}$ it would have been the dominant component   when
the universe was smaller by a factor larger than $\Omega_{DM}/\Omega_R\simeq 4\times 10^4\Omega_{DM}h^2$.

(6) Together we conclude that our universe has (approximately) $\Omega_{DE}\simeq 0.7,\Omega_{DM}\simeq 0.26,\Omega_B\simeq 0.04,\Omega_R\simeq 5\times 10^{-5}$. All known observations
are consistent with such an --- admittedly weird --- composition for the universe.

Using $\rho_{NR}\propto a^{-3}, \rho_R \propto a^{-4}$ and $\rho_{DE}$=constant   we can  write Eq.(\ref{frw}) in a convenient dimensionless form as
\begin{equation}
{1\over 2} \lb {dq\over d\tau}\rb^2 + V(q) = E 
\label{qomeganr}
\end{equation}
where $\tau = H_0t,a= a_0 q(\tau),\Omega_{\rm
NR} = \Omega_B + \Omega_{\rm DM}$ and 
\begin{equation}
V(q) = - {1\over 2} \left[ {\Omega_R\over q^2} + {\Omega_{\rm
NR}\over q} + \Omega_{DE} q^2\right]; \quad E={1\over 2} \lb 1-\Omega_{\rm tot}\rb.
\label{qveq}
\end{equation} 
This equation has the structure of the first integral for
motion of a particle with energy $E$ in a potential $V(q)$. 
For models with $\Omega  = \Omega_{\rm NR} + \Omega_{DE} =1$,
we can take $E=0$ so that $(dq/d\tau) = \sqrt{V(q)}$. Based on the observed composition of the universe, we can identify three distinct phases in 
the evolution of the universe when the temperature is less than about 100 GeV. At high redshifts (small $q$)
the universe is radiation dominated and $\dot q$ is independent 
of the other cosmological parameters. Then Eq.(\ref{qomeganr}) can be easily integrated
to give $a(t) \propto t^{1/2}$ and the temperature of the universe decreases as
$T\propto t^{-1/2}$. As the universe expands, a time will come
when ($t=t_{\rm eq}$,  $a=a_{\rm eq}$ and $z= z_{\rm eq}$, say)
the matter energy density will be comparable to radiation energy 
density. For the parameters described above, $(1+z_{eq})=\Omega_{NR}/\Omega_R\simeq 4\times 10^4\Omega_{DM}h^2$. At lower redshifts, matter will dominate over radiation and we will
have $a\propto t^{2/3}$ until fairly late when the 
dark energy density will dominate over
non relativistic matter.
This occurs at a redshift of $z_{\rm DE}$ where $(1+ z_{\rm DE}) =( \Omega_{\rm DE}/\Omega_{\rm NR})^{1/3}$.
For $\Omega_{\rm DE} \approx 0.7, \Omega_{\rm NR} \approx 0.3$, this occurs at $z_{\rm DE}\approx 0.33$.
In this phase, the velocity
$\dot q$ changes from being a decreasing function to an increasing function leading to 
an accelerating universe (see Fig.\ref{fig:tptrc}). In addition to these, we believe that the universe probably
went through a rapidly expanding, inflationary, phase very early when $T\approx 10^{14}$ GeV;
we will say more about this in Section \ref{sec:inflation}. (For a textbook description of these and related issues, see e.g. Ref.~\refcite{tpsfuv3}.)

\section{Growth of structures in the universe} 
 
 Having discussed the dynamics of the smooth universe, let us turn our attention to the formation
 of structures. In the conventional paradigm for the formation of structures in the universe, some
  mechanism is invoked to generate small perturbations in the energy density in the
  very early phase of the universe. These perturbations then grow due to gravitational
  instability and eventually form the  structures which we see today.
  Such a scenario is  constrained most severely by CMBR observations at $z\approx 10^3$. Since the
  perturbations in CMBR are observed to be small ($10^{-5} - 10^{-4}$ depending
  on the angular scale), it follows that the energy density perturbations were small compared
  to unity at the redshift of $z\approx 10^3$.

  The central quantity one uses to describe
  the growth of structures is 
   the {\it density contrast}
     defined as $\delta (t, {\bf x}) = [\rho(t,{\bf x}) - \rho_{\rm bg}(t)]/\rho_{\rm bg}(t)$ which characterizes 
     the fractional change in the energy density compared to the background. 
     Since one is 
     often interested in the statistical description of structures in the universe, it is conventional
     to assume that $\delta$ (and other related quantities) are  elements of a statistical ensemble.
     Many popular models of structure formation suggest that the initial density perturbations
     in the early universe can be represented as a Gaussian random variable with zero mean
      and a given initial power spectrum. The latter quantity is defined through
     the relation $P(t, k) = <|\delta_k(t)|^2>$ where $\delta_{\bf k}$ is the Fourier transform
     of $\delta(t,{\bf x})$ and $< ... >$ indicates  averaging over the 
     ensemble. The two-point correlation function
     $\xi(t,x)$ of the density distribution is defined as the Fourier transform of $P(t,{\bf k})$ over ${\bf k}$.
     
     When the $\delta \ll 1$, its evolution can be studied by linear perturbation
     theory and 
     each of the spatial Fourier modes $\delta_{\bf k}(t)$ will grow independently. Then the power 
 spectra $P(k,t) = <|\delta_{\bf k} (t)|^2>$  at two different times
 in the linear regime are related by
$ P(k,t_f)=\mathcal{F}^2(k,t_f,t_i, {\rm bg})P(k,t_i)$ 
 where $\mathcal{F}$  (called transfer function) depends only on the  parameters of the background universe  (denoted generically as ``bg'')  but
{\it not} on the initial power spectrum. 
The  form of  $\mathcal{F}$ is essentially decided by two factors: (i)
The relative magnitudes of the proper wavelength of perturbation $\lambda_{\rm prop} (t) \propto a(t)$ and the Hubble radius $d_H(t) \equiv H^{-1}(t)= (\dot a/a)^{-1}$
and (ii) whether the universe is radiation dominated or matter dominated.
At sufficiently early epochs, the universe will be radiation dominated and 
 the proper wavelength $\lambda_{\rm prop} (t) \propto a  \propto t^{1/2}$
 will be larger than
 $d_H(t) \propto t$. The density contrast of such modes, which are bigger than the 
Hubble radius, will grow\cite{tpsfuv3} as $ a^2$ until $\lambda_{\rm prop} = d_H(t)$.
(See the footnote on page \pageref{page:footnote}.) 
When this occurs, the perturbation at a given wavelength is said to
enter the Hubble radius.
If $\lambda_{\rm prop} < d_H$ and the universe is 
radiation dominated, the matter perturbation does not grow significantly and 
increases at best only logarithmically.\cite{tpsfuv3,meszaros} Later on, when the universe 
becomes matter dominated for $t>t_{\rm eq}$, the perturbations again
begin to grow. (Some of these details depend on the gauge chosen for describing the physics but, of course,
the final observable results are gauge independent; we shall not worry about this feature in this article.)
 
 It follows from this description that modes with wavelengths greater
than $d_{\rm eq} \equiv d_H (t_{\rm eq})$ --- which enter the Hubble
radius only in the matter dominated epoch --- continue to grow at all 
times;  modes with wavelengths smaller than $d_{\rm eq} $ suffer
lack of growth (in comparison with  longer wavelength modes) during the period
$t_{\rm enter} < t < t_{\rm eq}$.                        
This fact   distorts  the shape of the 
primordial spectrum by suppressing the growth of small wavelength modes
(with $k > k_{\rm eq}=2\pi/d_{\rm eq}$ that enter the Hubble radius in the radiation dominated phase)
in comparison with longer ones, with the transition occurring at the wave number $k_{eq}$
corresponding to the length scale 
$
d_{\rm eq} = d_H(z_{\rm eq})=  (2\pi/k_{\rm eq}) \approx 13 (\Omega_{\rm DM} h^2)^{-1}  {\rm Mpc}
$.
Very roughly, the shape of $\mathcal{F}^2(k)$ can be characterized by the behaviour
$\mathcal{F}^2(k) \propto k^{-4}$ for $k > k_{\rm eq}$ and $\mathcal{F}^2\approx 1$ for
$k< k_{\rm eq}$. 
The spectrum at wavelengths $\lambda \gg d_{\rm eq} $ is undistorted by
the evolution since $\mathcal{F}^2$ is essentially unity at these scales.   
 
We will see in the next section that inflationary models generate an initial power spectrum
of the form $P(k) \propto k$.  The evolution described above will distort it to the form
$P(k) \propto k^{-3}$ for $k > k_{\rm eq}$ and leave it undistorted 
with $P\propto k$ for $k< k_{\rm eq}$.
The power per logarithmic band in the wavenumber, $\Delta^2 \propto k^3 P(k)$, is approximately constant for $k>k_{\rm eq}$ (actually increasing as $\ln k$ because of the logarithmic growth
in the radiation dominated phase) and decreases as $\Delta^2 \propto k^4 \propto \lambda^{-4}$ at large wavelengths. It follows that $\Delta^2$ is a monotonically decreasing function of the wavelength with more power at small length scales.

 When $\delta_k \approx 1$, linear perturbation theory breaks down at the
 spatial scale corresponding to $\lambda= 2\pi/k$. Since there is more power at small scales, smaller scales go non-linear first and structure forms hierarchically. (Observations suggest that,
   in today's universe scales smaller than about $8 h^{-1} $ Mpc are non-linear; see Fig.\ref{fig:obs})
    As the universe expands, the over-dense region will expand more slowly compared to the background, 
 will reach a maximum radius, contract and virialize to form a bound nonlinear 
 halo of dark matter. The baryons in the halo will cool and undergo collapse
 in a fairly complex manner because of gas dynamical processes. 
 It seems unlikely that the baryonic collapse and galaxy formation can be understood
 by analytic approximations; one needs to do high resolution computer simulations
 to make any progress.\cite{baryonsimulations} 
 
 The non linear evolution of the  \textit{dark matter halos} is somewhat different
 and worth mentioning because it contains the  fascinating physics  of statistical mechanics
 of self gravitating systems.\cite{smofgs} 
 The standard instability of gravitating systems in a  \textit{static} background is 
 moderated  by  the presence of  a background expansion and it is possible to understand
 various features of nonlinear evolution of dark matter halos using different  analytic
 approximations.\cite{nlapprox} Among these, the existence of certain nonlinear scaling relations
 --- which allows one to compute nonlinear power spectrum from linear power spectrum by a
 nonlocal scaling relation --- seems to be most intriguing\cite{nsr}. If $\bar\xi(x,t)$ is the mean correlation function of dark matter particles and $\bar\xi_L(x,t)$ is the same quantity
 computed in the linear approximation, then, it turns out that $\bar\xi(x,t)$ can be expressed
 as a \textit{universal} function of $\bar\xi_L(x,t)$ in the form $\bar\xi(x,t)=U[\bar\xi_L(l,t)]$
 where $x=l[1+U[\bar\xi_L(l,t)]]^{-1/3}$. Incredibly enough, the form of $U$ can be determined by theory\cite{nsrtheory} and thus allows one to understand several aspects of nonlinear clustering
 analytically. This topic has interesting connections with renormalisation group theory, fluid turbulence etc. and deserves the attention of
 wider community of physicists.

\section{Inflation and generation of initial perturbations}\label{sec:inflation}

We saw that the two length scales which determine the evolution of perturbations are 
the Hubble radius $d_H(t)
\equiv (\dot a / a)^{-1}$
and $\lambda(t)\equiv \lambda_0 a(t)$. Using their definitions and Eq.(\ref{frw}), it is easy to show that
 if $\rho>0,p>0$, then
$\lambda(t)>d_H(t)$ for sufficiently small $t$.

This  result  leads to a major  difficulty  in  conventional 
cosmology.  Normal physical processes can act 
coherently only over length scales smaller than the Hubble radius.   
Thus 
any physical process leading to density perturbations  at 
some  early epoch, $t=t_i$, could only have operated at scales  smaller 
than $ d_H(t_i)$.  But most of the  relevant  astrophysical  scales 
(corresponding  to  clusters, groups, galaxies, etc.)  were  much 
bigger  than $d_H(t)$ at sufficiently early epochs.  
Therefore,   it is difficult to understand how  any  physical 
process operating in the early universe could have led to
the seed perturbations  in the  early 
universe.

One way of tacking this difficulty is to arrange matters such that  we have $\lambda(t)<d_H(t)$
at
sufficiently small $t$. Since we cannot do this in any model which has
both $\rho>0,p>0$ we need to invoke some exotic physics to get around this difficulty.
The standard procedure is to make $a(t)$ increase rapidly with $t$ (for example, exponentially
or as $a\propto t^n$ with $n\gg 1$, which requires $p<0$)  for a brief period of time. 
Such a rapid growth is called
``inflation" and in
 conventional models of inflation,\cite{inflation} the energy density during the inflationary phase is provided by a scalar field with a potential $V(\phi)$. If  the potential energy dominates over the kinetic energy, such a scalar field can act like an ideal fluid with the equation of state $p = - \rho$ and lead to $a(t)\propto e^{Ht}$ during inflation. 
 Fig. (\ref{fig:tpplumian}) shows the behaviour of the Hubble radius and the wavelength $\lambda(t)$ of a generic perturbation (line AB) for a universe which underwent exponential inflation. In such a universe, it is possible for \textit{quantum} fluctuations of the scalar field at A (when the perturbation scale leaves the Hubble radius) to manifest as \textit{classical} perturbations at B (when the perturbation enters the Hubble radius).
We will now briefly discuss these processes.

Consider a scalar field $\phi(t, {\bf x})$ which is nearly homogeneous in the sense that
we can write $\phi(t, {\bf x}) = \phi(t) + \delta \phi(t,{\bf x})$ with $\delta \phi \ll \phi$.
Let us first ignore
the fluctuations and consider how one can use the mean value to drive a
rapid expansion of the universe. 
 The Einstein's equation (for $k=0$) with the field $\phi(t)$ as the source can be written in the form 
\begin{equation} \frac{\dot a^2}{a^2} = H^2(t)  = \frac{1}{3 M^2_{\rm Pl}}  \left[  \frac{1}{2} \dot \phi^2 + V(\phi) \right] 
\label{qthreedotfive}
\end{equation}
where $V(\phi)$ is the potential for the scalar field and $M_{\rm Pl} \equiv (8\pi G)^{-1/2} \approx 2.4 \times 10^{18}$ GeV in units  with $\hbar = c = 1$. 
Further, the equation of motion for the scalar field in an expanding universe reduces to  
\begin{equation} \ddot \phi + 3H\dot \phi = - \frac{dV}{d\phi}
\label{qthreedotsix}
\end{equation}
The solutions  of Eqs.~(\ref{qthreedotfive}), (\ref{qthreedotsix}) giving $a(t)$ and
$\phi(t)$  will depend critically on the form of $V(\phi)$ as well
as the initial conditions. Among these solutions, there exists a subset in 
which $a(t)$ is a rapidly growing function of $t$, either exponentially
or as a power law $a(t) \propto t^n$ with an arbitrarily large value of $n$.
It is fairly easy to verify that the solutions to 
Eqs.~(\ref{qthreedotfive}), (\ref{qthreedotsix}) can be expressed in the form 
\begin{equation}
 V(t) = 3H^2 M_{Pl}^2\left[ 1+ \frac{\dot H}{3H^2}\right];\quad
 \phi(t) = \int dt [ -2\dot H M_{Pl}^2]^{1/2}
\label{qmainphi}
\end{equation}
Equation (\ref{qmainphi}) completely solves the (reverse) problem
of finding a potential $V(\phi)$ which will lead to a given
$a(t)$. For example, 
power law expansion of the universe [$a\propto t^n$] can be generated by
 using a potential $V\propto \exp[-\sqrt{(2/n)}(\phi/M_{Pl})]$. 

A more generic way of achieving this is through potentials which allow what is known as {\it slow roll-over}.
 Such potentials have a gently decreasing form for 
$V(\phi)$ for a range of values for $\phi$ allowing  $\phi(t)$ to evolve 
very slowly. Assuming a sufficiently slow evolution  of $\phi(t)$ 
we can ignore: (i) the $\ddot \phi$ term in equation \eq{qthreedotsix} and  (ii) the kinetic energy term $\dot \phi^2$ in comparison with the potential
energy $V(\phi)$ in  \eq{qthreedotfive}. In this limit, 
\eq{qthreedotfive},\eq{qthreedotsix} become
\begin{equation} H^2 \simeq \fra{V(\phi)}{3M_{\rm Pl}^2}; \qquad 3 H \dot \phi \simeq - V'(\phi)
\label{qthreedoteight}
\end{equation}
The validity of slow roll over approximation thus
requires the following two parameters to be sufficiently small:
\begin{equation} \epsilon (\phi) = \fra{M_{\rm Pl}^2}{2} \frab{V'}{V}^2; \qquad \eta(\phi) =  M_{\rm Pl}^2\, \fra{V''}{V} 
\label{etaepsilon}
\end{equation}
 The end point for inflation
can be taken to be the epoch at which $\epsilon$ becomes comparable to unity.
If the slow roll-over approximation is valid until a time
$t= t_{\rm end}$, the amount of inflation can be characterized by the ratio
$a(t_{\rm end})/a(t)$. If $N(t) \equiv \ln [a(t_{\rm end}/a(t)]$, then  \eq{qthreedoteight} gives
\begin{equation} 
N \equiv  \ln \fra{a(t_{\rm end})}{a(t)} = \int_t^{t_{\rm end}} H\, dt \simeq \fra{1}{M_{\rm Pl}^2} \int_{\phi_{\rm end}}^\phi \fra{V}{V'}\, d\phi
\label{defN}
\end{equation}
This provides a general procedure for  quantifying the rapid growth of $a(t)$ arising from a given potential. 

Let us next consider the spectrum of density perturbations which are generated from the quantum fluctuations of the scalar field.\cite{genofpert}
This requires the study of 
quantum field theory in a time dependent background which is non-trivial. There are 
several conceptual issues (closely related to the issue of 
general covariance of quantum field theory and the  particle concept\cite{probe}) in obtaining a c-number density perturbation from inherently quantum fluctuations. We shall not discuss these issues and will adopt a
heuristic approach, as follows: 

In the
deSitter spacetime with $a(t) \propto \exp(Ht)$, there is a horizon in the spacetime
and associated temperature $T=(H/2\pi)$  --- just as in the case of black holes.\cite{ghds} 
Hence the scalar field will have an intrinsic rms fluctuation
$\delta\phi \approx T = (H/2\pi)$ in the deSitter spacetime at the scale of the Hubble radius. This will cause a time shift $\delta t \approx \delta \phi/\dot \phi$ in the 
evolution of the field between  patches of the universe of size about $H^{-1}$. This, in turn, will lead to 
an rms fluctuation $\Delta=(k^3P)^{1/2}$ of amplitude $\delta a/a=(\dot a/a)\delta t\approx H^2/(2\pi \dot \phi)$ at the Hubble scale.
Since the wavelength of the perturbation is equal to Hubble radius at A (see Fig. (\ref{fig:tpplumian})), we conclude that the
rms amplitude of 
 the perturbation when it leaves the Hubble radius is: $\Delta_A\approx H^2/(2\pi \dot \phi)$. Between A and B (in Fig. (\ref{fig:tpplumian})) the wavelength of the perturbation is bigger than the 
Hubble radius and one can show\label{page:footnote}\footnote{For a single component universe with $p=w\rho \propto a^{-3(1+w)}$, 
we have from Eq.~(\ref{frw}), the result $\ddot a = - (4\pi G/3)
(1+3w) \rho a$. Perturbing this relation to $a \to a+\delta a $ and using $a=(t/t_0)^{2/(3+3w)}$
we find that $\delta a $ satisfies the equation $t^2 \ddot{\delta a} = m\delta a$ with
$m=(2/9) (1+3w)(2+3w)(1+w)^{-2}$. This has power law solutions $\delta a \propto t^p$ with
$p(p-1) = m$. The growing mode corresponds to the density contrast $\delta \propto (\delta a/a)$
which is easily shown to vary as $\delta \propto (\rho a^2)^{-1}$. In the inflationary, phase, $\rho$=const., $\delta\propto a^{-2}$; in the radiation dominated
phase, $\rho \propto a^{-4},\delta \propto a^2$. The result follows from these scalings. } that $\Delta(at\ A)\approx\Delta(at\ B)$
giving $\Delta(at\ B)\approx H^2/(2\pi \dot \phi)$. Since this is independent of $k$, it follows that all perturbations enter the
Hubble radius with constant power per decade. That is $\Delta^2(k,a)\propto k^3P(k,a)$ is independent of $k$ when evaluated at
$a=a_{enter}(k)$ for the relevant mode. 

From this, it follows that $P(k,a)\propto k$ at constant $a$. To see this, note
that if $P\propto k^n$, then the power per logarithmic band of wave numbers
is $\Delta^2\propto k^3P(k)\propto k^{(n+3)}$. Further, when the wavelength of the mode is larger than the Hubble radius,  during the radiation dominated phase,
the perturbation 
grows (see footnote)
as $\delta \propto a^2$ making $\Delta^2\propto a^4 k^{(n+3)}$.  The epoch $a_{enter}$ at which a mode enters the Hubble
radius
is determined by the relation $2\pi a_{enter}/k=d_H$. Using $d_H\propto
t\propto a^2$ in the radiation dominated phase, we get $a_{enter}\propto k^{-1}$
so that
\begin{equation}
\Delta^2(k,a_{enter})\propto a_{enter}^4 k^{(n+3)}\propto k^{(n-1)}
\end{equation}
So if the power $\Delta^2 \propto k^3 P$ per octave in $k$ is independent of scale $k$, at the
time of entering the Hubble radius, then $n=1$. 
In fact, a \emph{prediction} that the initial fluctuation spectrum will have a power spectrum
$P=Ak^n$ with $n=1$ was  made by
 Harrison 
and Zeldovich,\cite{zeldovich72} years before inflationary paradigm,
  based on  general arguments of scale invariance.
 Inflation is one possible mechanism for generating such  scale invariant perturbations.

As an example, consider the case of $V(\phi) = \lambda \phi^4$, for which Eq.(\ref{defN}) gives
$N = H^2/2\lambda \phi^2$ and the amplitude of the perturbations at Hubble scale is:
\begin{equation}
\Delta \simeq \frac{H^2}{\dot \phi} \simeq \frac{3H^3}{V'} \simeq \lambda^{1/2} N^{3/2}
\end{equation}
If we want inflation to last for reasonable amount of time ($N \gtrsim 60$, say)
and $\Delta \approx 10^{-5}$ (as determined from CMBR temperature anisotropies;
see Section \ref{sec:tempcmbr}), then we require $\lambda \lesssim 10^{-15}$. This has been
a serious problem in virtually any reasonable model of inflation: \textit{The parameters
in the potential need
to be extremely fine tuned to match observations.}

It is not possible to obtain $n$ strictly equal
to unity in realistic models, since scale invariance is always broken at some level. 
In a wide class of inflationary models this deviation is given by
$(1-n) \approx 6\epsilon - 2\eta$ (where $\epsilon$ and $\eta$ are defined by 
Eq.~(\ref{etaepsilon})); this deviation $(1-n)$ is obviously small when the slow roll over approximation
($\epsilon \ll 1, \eta \ll 1$) holds.

The same mechanism that produces density perturbations (which are scalar) will also
produce gravitational wave (tensor) perturbations of some magnitude $\delta_{grav}$. 
Since both the scalar and tensor perturbations arise from the same mechanism,
one can relate the amplitudes of these two and show that $(\delta_{grav}/\delta)^2
\approx 12.4\epsilon$; clearly,  the tensor perturbations
are small compared to the scalar perturbations. Further, for generic inflationary potentials, $|\eta| \approx |\epsilon|$
so that $(1-n) \approx 4 \epsilon$ giving $(\delta_{grav}/\delta)^2 \approx \mathcal{O} (3)(1-n)$.
This is a relation between three quantities all of which are (in principle) directly observable
and hence it can provide a test of the underlying model if and when we detect the stochastic gravitational wave background.

Finally, we mention the possibility that inflationary
regime might act as a magnifying glass and bring the transplanckian regime of physics within the scope of direct
observations.\cite{transplanck}. To see how this could be possible, note that a scale $\lambda_0$ today would have been
$\lambda_f\equiv\lambda_0(a_f/a_0)=\lambda_0(T_0/T_f) = 3\lambda_0 \times 10^{-27}$ at the end of inflation and $\lambda_i=\lambda_f\exp(-N) \simeq \lambda_f e^{-70}$
at the beginning of inflation,  for  typical numbers used in the inflationary scenario. This gives $\lambda_i\approx
3L_P(\lambda_0/1 $ Mpc) showing that most of the astrophysically relevant scales were smaller than Planck length, $L_P \equiv (G\hbar/c^3)^{1/2} \simeq 10^{-33}$ cm,  during the inflation! Phenomenological models which make specific predictions regarding transplanckian physics
(like dispersion relations,\cite{dispersion} for example) can then be tested using the signature they leave on the
pattern of density perturbations which are generated.

\section{Temperature anisotropies of the CMBR}\label{sec:tempcmbr}

When the universe cools through $T\approx1$
eV, the electrons combine with nuclei forming neutral atoms. This `re'combination takes place
at a redshift of about $z\approx10^3$ over a redshift interval $\Delta z=80$. Once neutral atoms form, the photons decouple from matter and propagate freely from $z=10^3$ to $z=0$. This CMB
radiation, therefore, contains fossilized signature of the conditions of the universe at $z=10^3$
and has been an invaluable source of information.  

If physical process has led to inhomogeneities in  the $z=10^3$
   spatial surface, then these inhomogeneities will appear as temperature anisotropies
   $(\Delta T/T) \equiv S(\theta,\phi)$ of the CMBR in the sky today where $(\theta,\phi)$ denotes two angles in the sky. 
   It is convenient to expand this quantity in spherical harmonics as $S(\theta,\phi) = \sum a_{lm} Y_{lm}(\theta,\phi)$.
 If ${\bf n}$ and ${\bf m}$ are two directions in the sky with an angle $\alpha$ between them, the two-point correlation function of the temperature fluctuations in the sky can be 
 expressed in the form
\begin{equation}
{\mathcal C} (\alpha) \equiv \langle S(\textbf{n}) S(\textbf{m})\rangle 
= \sum_l {(2l+1)\over 4\pi} C_l P_l(\cos \alpha); \quad  C_l  = \langle|a_{lm}|^2\rangle
\label{qcalpha}
\end{equation}  
 Roughly speaking, $l\propto \theta^{-1}$ and we can
think of the $(\theta, l)$ pair as analogue of $({\bf x}, {\bf k})$ variables
in 3-D. 

The primary anisotropies of the CMBR  can be thought of as arising from  three different sources
(even though such a separation is gauge dependent).
(i) The first is the gravitational potential fluctuations at the last
scattering surface (LSS) which will contribute an anisotropy $(\Delta T/T)_\phi^2
\propto k^3P_\phi(k)$ where $P_\phi(k)\propto P(k)/k^4$ is the 
power spectrum of gravitational potential $\phi$. 
(The gravitational potential satisfies $\nabla^2 \phi \propto \delta$ which becomes
$k^2 \phi_k \propto \delta_k$ in Fourier space; so $P_\phi \equiv \langle |\phi_k|^2 \rangle
\propto k^{-4} \langle |\delta_k|^2 \rangle \propto P(k) /k^4$.)
This anisotropy arises
because photons climbing out of deeper gravitational wells lose
more energy on the average.  
(ii) The second source is the Doppler shift of the frequency of the photons
when they are last scattered by moving electrons on the LSS.
This is proportional to $(\Delta T/T)_D^2\propto k^3 P_v$ where 
$P_v(k)\propto P/k^2$ is the power spectrum of the velocity field.
(The velocity field is given by ${\bf v} \simeq {\bf g} t \propto t \nabla \phi$ 
so that, in Fourier space, $v_k \sim k \phi_k $ and $P_v=|v_k|^2 \propto k^2 P_\phi\propto k^{-2} P$.)
(iii) Finally, we also need to take into account the intrinsic 
fluctuations of the radiation field on the LSS. In the case of adiabatic 
fluctuations, these will be proportional to the density fluctuations of
matter on the LSS and hence will vary as $(\Delta T/T)_{\rm int}^2\propto k^3P(k)$.
Of these, the velocity field and the density field (leading to the 
Doppler anisotropy and intrinsic anisotropy described in (ii) and (iii) above)
will oscillate at scales smaller than the Hubble radius at 
the time of decoupling
since pressure support due to baryons will be effective at small scales.
At large scales, for a scale invariant spectrum with $P(k)\propto k$, we get:
\begin{equation}
\left({\Delta T\over T}\right)_\phi^2\propto \ {\rm const}; \quad 
\left({\Delta T\over T}\right)_D^2\propto k^2\propto \theta^{-2}; \quad 
\left({\Delta T\over T}\right)_{\rm int}^2\propto k^4\propto \theta^{-4}
\end{equation}
where $\theta\propto \lambda \propto k^{-1}$ is the angular scale
over which the anisotropy is measured. The fluctuations due
to gravitational potential dominate at large scales while 
 the sum of intrinsic and Doppler anisotropies
will dominate at small scales. Since the latter two 
are oscillatory, we will expect an oscillatory behaviour in the 
temperature anisotropies at small angular scales. 
The typical value
for the peaks of the oscillation are at about 0.3 to 0.5 degrees depending
on the details of the model.

 The above analysis 
is valid if recombination was instantaneous; but in reality the thickness
of the recombination epoch is about $\Delta z\simeq 80$. Further, the coupling between the photons and baryons is not
completely `tight'. It can be shown that\cite{tpsfuv3} these features will heavily damp the
 anisotropies  at  angular scales smaller than about
0.1 degree.

The fact that  different processes contribute to the 
 structure of angular anisotropies makes CMBR  a valuable
tool for extracting cosmological information. To begin with, the 
anisotropy at very large scales directly probes modes which are 
bigger than the Hubble radius at the time of decoupling
and  allows us to directly determine the primordial spectrum.
The CMBR observations are \textit{consistent} with the inflationary model for the generation of perturbations leading to $P=Ak^n$ and gives $A\simeq (28.3 h^{-1} Mpc)^4$ and $n=0.97\pm0.023$. (The first results\cite{cobeanaly} were from COBE and later results, especially from
WMAP, have reconfirmed\cite{cmbr} them with far greater accuracy).
 
 \begin{figure}[ht]
\begin{center}
\includegraphics[angle=-90,scale=.5]{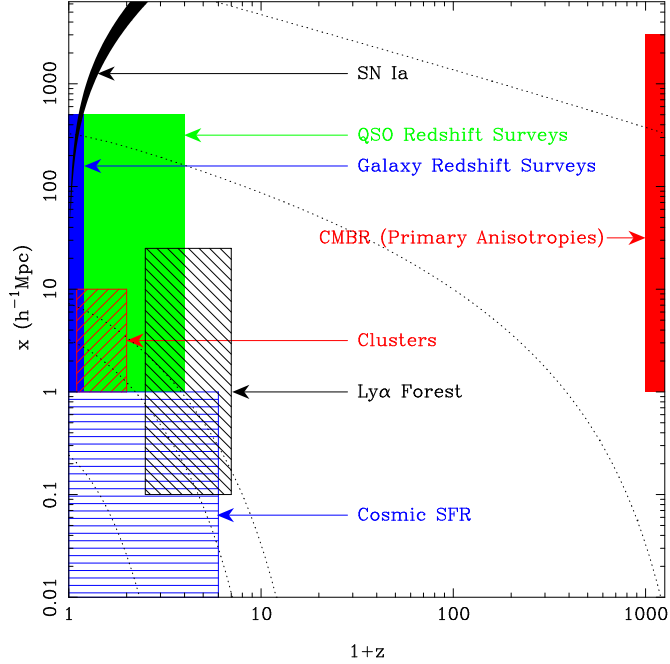}
\end{center}
\caption{Different observations of cosmological significance and the length scales and redshift ranges probed by them. The broken (thin) lines in the figure are contours of $\Delta(k=2\pi/x,a)=(5,1.69,1,10^{-2},10^{-5})$ from bottom to top.
Figure courtesy:T.RoyChoudhury
}
\label{fig:obs}
\end{figure}

As we move to smaller scales we are probing the behaviour of baryonic gas coupled to the
photons. The pressure support of the gas leads to modulated acoustic oscillations
with a characteristic wavelength at the $z=10^3 $ surface. Regions of high and low
baryonic density contrast will lead to anisotropies in the temperature with the same
characteristic wavelength (which acts as a standard ruler) leading to a series of peaks in the temperature anisotropy
that have been detected.  
 The angles subtended by  these
acoustic peaks  will depend on the geometry of the universe
and
 provides a  reliable procedure for estimating the 
cosmological parameters. 
Detailed computations\cite{tpsfuv3} show that: (i) The multipole index $l$ corresponding to the first acoustic
  peak has a strong, easily observable, dependence on $\Omega_{\rm tot}$ and scales as $l_p \approx 220 \Omega_{\rm tot}^{-1/2}$  if there is \textit{no}
  dark energy and $\Omega_{tot}=\Omega_{NR}$. (ii) But if both non-relativistic matter and dark energy is present, with 
  $\Omega_{\rm NR} + \Omega_{DE} =1$ and
  $0.1 \lesssim \Omega_{\rm NR} \lesssim 1$, then the peak has only a very weak dependence on $\Omega_{NR}$ and
  $l_p \approx 220\Omega_{NR}^{0.1}$. Thus the observed location of the peak (which is around $l\sim 220)$
  can be used to infer that  $\Omega_{\rm tot} \simeq 1$. More precisely, the current observations show that $0.98\lesssim\Omega_{\rm tot}\lesssim1.08$; combining with $h>0.5$, this result implies the existence of
  dark energy.

   The heights of acoustic peaks also contain important information.
In particular, the height of the first acoustic peak
relative to the second one depends sensitively on $\Omega_B$ and the current results are consistent with that obtained from big bang nucleosynthesis.\cite{baryon} 

Fig.\ref{fig:obs} summarises the different observations of cosmological significance and the range of length scales and redshift ranges
probed by them. The broken (thin) lines in the figure are contours of $\Delta(k=2\pi/x,a)=(5,1.69,1,10^{-2},10^{-5})$ from bottom to top. Clearly,
the regions where   $\Delta(k,a)>1$ corresponds to those in which nonlinear effects of structure formation is important. Most of the astrophysical observations --- large scale surveys of galaxies, clusters
and quasars,  observations of intergalactic medium (Ly-$\alpha$ forest), star formation rate (SFR), supernova (SN Ia) data etc. --- are confined to $0<z\lesssim7$ while CMBR allows probing the universe around $z=10^3$. Combining these allows one to use the long ``lever arm"
of 3 decades in redshift and thus constrain the parameters describing the universe effectively.

\section{The Dark Energy}

It is rather frustrating that we have no direct laboratory evidence for nearly 96\% of matter in the universe.
(Actually, since we do not quite understand the process of baryogenesis, we do not understand $\Omega_B$ either;
all we can \textit{theoretically} understand now is a universe filled entirely with radiation!). Assuming that particle physics models will eventually  (i) explain  $\Omega_B$ and $\Omega_{DM}$ (probably arising from the lightest
supersymmetric partner) as well as (ii) provide a viable model for inflation predicting correct value for $A$,
one is left with the problem of understanding $\Omega_{DE}$. While the issues (i) and (ii) are by no means trivial or satisfactorily addressed, the issue of dark energy is lot more perplexing,
thereby justifying the attention it has received recently.

\begin{figure}[ht]
\begin{center}
\includegraphics[angle=-90,scale=.5]{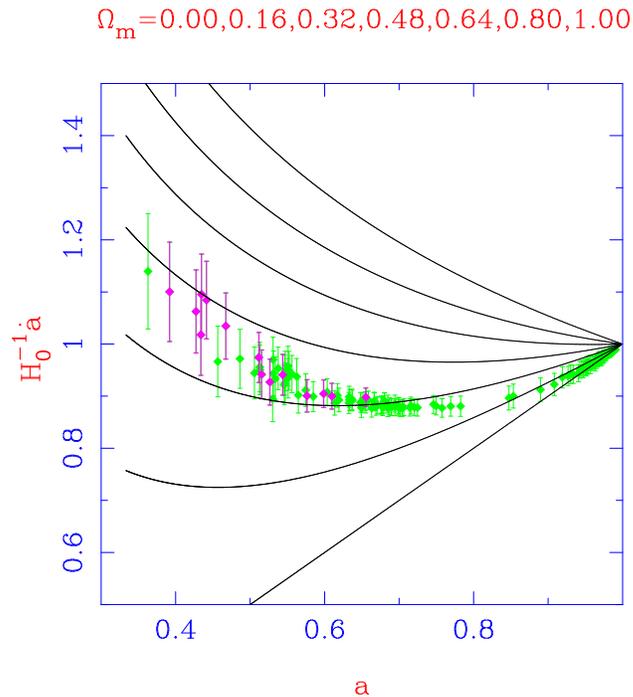}
\end{center}
\caption{The ``velocity'' $\dot a$  of the universe is plotted against the ``position'' $a$ in the form of a phase portrait. The different curves are for models parameterized by the value of  $\Omega_{DM}(=\Omega_m)$ keeping $\Omega_{tot}=1$. The top-most curve has $\Omega_m=1$ and the bottom-most curve has $\Omega_m=0$
and $\Omega_{DE}=1$. The in-between curves show  universes which were decelerating in the past and began to accelerate when the dark energy started dominating. The supernova data clearly favours such a model. 
}
\label{fig:tptrc}
\end{figure}

The key observational feature of dark energy is that --- treated as a fluid with a stress tensor $T^a_b={\rm dia} (\rho, -p, -p,-p)$ 
--- it has an equation of state $p=w\rho$ with $w \lesssim -0.8$ at the present epoch. 
The spatial part  ${\bf g}$  of the geodesic acceleration (which measures the 
  relative acceleration of two geodesics in the spacetime) satisfies an \textit{exact} equation
  in general relativity  given by
  $
  \nabla \cdot {\bf g} = - 4\pi G (\rho + 3p).
 $
  As long as $(\rho + 3p) > 0$, gravity remains attractive while $(\rho + 3p) <0$ can
  lead to repulsive gravitational effects. In other words, dark energy with sufficiently negative pressure will
  accelerate the expansion of the universe, once it starts dominating over the normal matter.  This is precisely what is established from the study of high redshift supernova, which can be used to determine the expansion
rate of the universe in the past.\cite{sn} 
Figure \ref{fig:tptrc} presents the supernova data as a phase portrait\cite{tptirthsn1} of the universe. It is  clear that the universe was decelerating at high redshifts and started accelerating when it was about two-third of the present size. 

The simplest model for  a fluid with negative pressure is the
cosmological constant\cite{cc} with $w=-1,\rho =-p=$ constant.
If the dark energy is indeed a cosmological constant, then it introduces a fundamental length scale in the theory $L_\Lambda\equiv H_\Lambda^{-1}$, related to the constant dark energy density $\rho_{DE}$ by 
$H_\Lambda^2\equiv (8\pi G\rho_{DE}/3)$.
In classical general relativity,
    based on the constants $G, c $ and $L_\Lambda$,  it
  is not possible to construct any dimensionless combination from these constants. But when one introduces the Planck constant, $\hbar$, it is  possible
  to form the dimensionless combination $H^2_\Lambda(G\hbar/c^3) \equiv  (L_P^2/L_\Lambda^2)$.
  Observations demand $(L_P^2/L_\Lambda^2) \lesssim 10^{-120}$
   requiring enormous fine tuning. What is more, the energy density of 
  normal matter and radiation  would have been higher in the past  while the energy density contributed by the  cosmological constant
  does not change.  Hence we need to adjust the energy densities
  of normal matter and cosmological constant in the early epoch very carefully so that
  $\rho_\Lambda\gtrsim \rho_{\rm NR}$ around the current epoch.
  Because of these conceptual problems associated with the cosmological constant, people have explored a large variety of alternative possibilities. Though none of them does any better than the cosmological constant, we will briefly describe them in view of the popularity these models enjoy.

  The most popular alternative to \cc\ uses a scalar field $\phi$ with a suitably chosen potential $V(\phi)$ so as to make the vacuum energy vary with time. The hope  is that, one can find a model in which the current value can be explained naturally without any fine tuning.
  We will discuss two possibilities based on the lagrangians:
    \begin{equation}
  L_{\rm quin} = \frac{1}{2} \partial_a \phi \partial^a \phi - V(\phi); \quad L_{\rm tach}
  = -V(\phi) [1-\partial_a\phi\partial^a\phi]^{1/2}
  \label{lquineq}
  \end{equation}
  Both these lagrangians involve one arbitrary function $V(\phi)$. The first one,
  $L_{\rm quin}$,  which is a natural generalization of the lagrangian for
  a non-relativistic particle, $L=(1/2)\dot q^2 -V(q)$, is usually called quintessence.\cite{phiindustry}
    When it acts as a source in Friedman universe,
   it is characterized by a time dependent
  $w(t) = (1-(2V/\dot\phi^2))(1+ (2V/\dot\phi^2))^{-1}$. 

The structure of the second lagrangian  in Eq.~(\ref{lquineq}) can be understood by an analogy  with a relativistic particle with   position
$q(t)$ and mass $m$ which  is described by the lagrangian $L = -m \sqrt{1-\dot q^2}$.  We can now
construct a field theory by upgrading $q(t)$ to a field $\phi$ and treating the mass parameter $m$ as a function of
$\phi$ [say, $V(\phi)$] thereby obtaining the second lagrangian in Eq.~(\ref{lquineq}). This provides a rich gamut of possibilities in the
context of cosmology.\cite{tptachyon,tachyon}
  This form of scalar field arises  in string theories\cite{asen} and
   is called a tachyonic scalar field.
   (The structure of this lagrangian is similar to those analyzed previously in a  class of models\cite{kessence}
   called {\it K-essence}.)
The stress tensor for the tachyonic scalar  field can be written 
as  the sum of a pressure less dust component and a cosmological constant.
This suggests a possibility\cite{tptachyon} of providing a unified description of both dark matter
and dark energy using the same scalar field.
(It is possible to construct more complicated scalar field lagrangians  with even $w<-1$ describing
   what is called {\it phantom} matter; there are also 
   alternatives to scalar field models, based on brane world scenarios. We shall not discuss either of these.)
  
   Since  the quintessence or the tachyonic field   has
   an undetermined function $V(\phi)$, it is possible to choose this function
  in order to produce a given $H(a)$.
  To see this explicitly, let
   us assume that the universe has two forms of energy density with $\rho(a) =\rho_{\rm known}
  (a) + \rho_\phi(a)$ where $\rho_{\rm known}(a)$ arises from any known forms of source 
  (matter, radiation, ...) and
  $\rho_\phi(a) $ is due to a scalar field.  
  Let us first consider quintessence. Here,  the potential is given implicitly by the form\cite{ellis,tptachyon}
  \begin{equation}
  V(a) = \frac{1}{16\pi G} H (1-Q)\left[6H + 2aH' - \frac{aH Q'}{1-Q}\right]
  \label{voft}
   \end{equation} 
    \begin{equation}
    \phi (a) =  \left[ \frac{1}{8\pi G}\right]^{1/2} \int \frac{da}{a}
     \left[ aQ' - (1-Q)\frac{d \ln H^2}{d\ln a}\right]^{1/2}
    \label{phioft}
    \end{equation} 
   where $Q (a) \equiv [8\pi G \rho_{\rm known}(a) / 3H^2(a)]$ and prime denotes differentiation with respect to $a$.
   Given any
   $H(a),Q(a)$, these equations determine $V(a)$ and $\phi(a)$ and thus the potential $V(\phi)$. 
   \textit{Every quintessence model studied in the literature can be obtained from these equations.}

 Similar results exists for the tachyonic scalar field as well.\cite{tptachyon}
  For example, given
  any $H(a)$, one can construct a tachyonic potential $V(\phi)$ which is consistent with it. The equations determining $V(\phi)$  are now given by:
  \begin{equation}
  \phi(a) = \int \frac{da}{aH} \left(\frac{aQ'}{3(1-Q)}
   -{2\over 3}{a H'\over H}\right)^{1/2}
  \label{finalone}
  \end{equation}
   \begin{equation}
   V = {3H^2 \over 8\pi G}(1-Q) \left( 1 + {2\over 3}{a H'\over H}-\frac{aQ'}{3(1-Q)}\right)^{1/2}
   \label{finaltwo}
   \end{equation}
   Again, Eqs.~(\ref{finalone}) and (\ref{finaltwo}) completely solve the problem. Given any
   $H(a)$, these equations determine $V(a)$ and $\phi(a)$ and thus the potential $V(\phi)$. 
    A wide variety of phenomenological models with time dependent
  \cc\ have been considered in the literature all of which can be 
   mapped to a 
  scalar field model with a suitable $V(\phi)$. 
  
  While the scalar field models enjoy considerable popularity (one reason being they are easy to construct!)
  they have not helped us to understand the nature of the dark energy
  at a deeper level because of several shortcomings:
  \begin{figure}[ht]
 \begin{center}
 \includegraphics[scale=0.5]{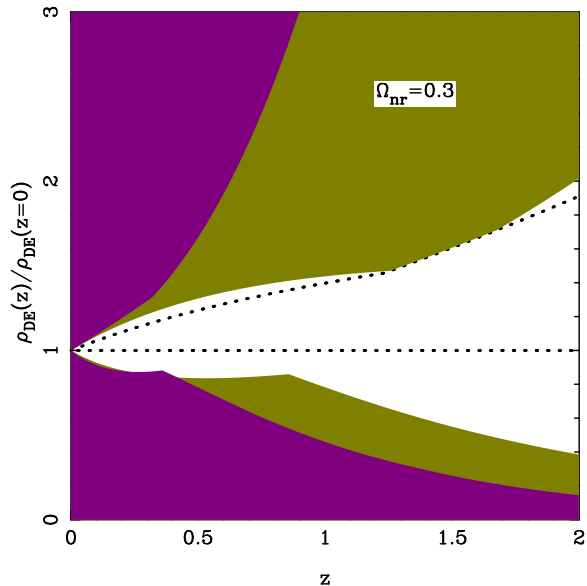}
 \end{center}
 \caption{Constraints on the possible variation of the dark energy density with redshift. The darker shaded region (magenta) is excluded by SN observations while the lighter shaded region (green) is excluded by WMAP observations. It is obvious that WMAP puts stronger constraints on the possible
 variations of dark energy density. The cosmological constant corresponds to the horizontal line
 at unity. The region between the dotted lines has $w>-1$ at all epochs.
  }
 \label{fig:bjp2ps}
 \end{figure}  
  (1) They completely lack predictive power. As explicitly demonstrated above, virtually every form of $a(t)$ can be modeled by a suitable ``designer" $V(\phi)$.
  (2) These models are  degenerate in another sense. Even when $w(a)$ is known/specified, it is not possible to proceed further and determine
  the nature of the scalar field lagrangian. The explicit examples given above show that there
  are {\em at least} two different forms of scalar field lagrangians (corresponding to
  the quintessence or the tachyonic field) which could lead to
  the same $w(a)$. (See Ref.~\refcite{tptirthsn1} for an explicit example of such a construction.)
  (3) All the scalar field potentials require fine tuning of the parameters in order to be viable. This is obvious in the quintessence models in which adding a constant to the potential is the same as invoking a \cc. So to make the quintessence models work, \textit{we first need to assume the \cc\ is zero!}
  (4) By and large, the potentials  used in the literature have no natural field theoretical justification. All of them are non-renormalisable in the conventional sense and have to be interpreted as a low energy effective potential in an ad-hoc manner.
  
  One key difference between \cc\ and scalar field models is that the latter lead to a $w(a)$ which varies with time. If observations have demanded this, or even if observations have ruled out $w=-1$ at the present epoch,
  then one would have been forced to take alternative models seriously. However, all available observations are consistent with \cc\ ($w=-1$) and --- in fact --- the possible variation of $w$ is strongly constrained\cite{jbp} as shown in Figure \ref{fig:bjp2ps}.

Given this situation, we shall  take a closer look at the \cc\ as the source of dark energy in the universe.
 
 \section{...For the Snark was a Boojum, you see }
 
\begin{figure}
  \includegraphics[angle=-90,scale=0.45]{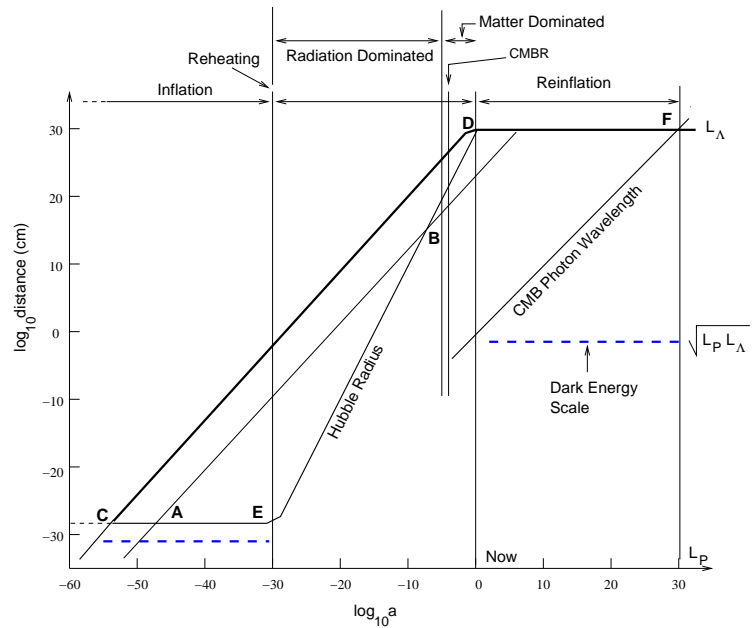}
\caption{The geometrical structure of a universe with two length scales $L_P$ and $L_\Lambda$ corresponding to the Planck length and the cosmological constant. See text for detailed description of the figure.
}
\label{fig:tpplumian}
  \end{figure}

 If we assume that the dark energy in the universe is due to a \cc\, then we are introducing a \textit{second} length scale, $L_\Lambda=H_\Lambda^{-1}$, into the theory  (in addition to the Planck length $L_P$) such that 
$ (L_P/L_\Lambda)\approx
 10^{-60}$. Such a universe will be asymptotically deSitter with $a(t)\propto \exp (t/L_\Lambda) $ at late times.
 Figure \ref{fig:tpplumian} summarizes
  several peculiar features of such a universe.\cite{plumian,bjorken}

  Using the 
 the Hubble radius $d_H\equiv (\dot a/a)^{-1}$, we can distinguish between three different phases of such a universe. The first phase is when the universe went through a inflationary expansion with $d_H=$ constant; the second phase is the radiation/matter dominated phase in which most of the standard cosmology operates and $d_H\propto t$ increases monotonically; the third phase is that of re-inflation (or accelerated expansion) governed by the cosmological constant in which $d_H$ is again a constant. The first and last phases are time translation invariant;
 that is, $t\to t+$ constant is an (approximate) invariance for the universe in these two phases. The universe satisfies the perfect cosmological principle and is in steady state during these phases!
In fact, one can easily imagine a scenario in which the two deSitter phases (first and last) are of arbitrarily long duration.\cite{plumian} If  $\Omega_\Lambda\approx 0.7, \Omega_{DM}\approx 0.3$ the final deSitter phase \textit{does} last forever; as regards the inflationary phase, one can view it as lasting for arbitrarily long duration. 
 
    Given the two length scales $L_P$ and $L_\Lambda$, one can construct two energy densities
 $\rho_P=1/L_P^4$ and $\rho_\Lambda=1/L_\Lambda^4$ in natural units ($c=\hbar=1$). The first is, of course, the Planck energy density while the second one also has a natural interpretation. The universe which is asymptotically deSitter has a horizon and associated thermodynamics\cite{ghds} with a  temperature
 $T=H_\Lambda/2\pi$ and the corresponding thermal energy density $\rho_{thermal}\propto T^4\propto 1/L_\Lambda^4=
 \rho_\Lambda$. Thus $L_P$ determines the \textit{highest} possible energy density in the universe while $L_\Lambda$
 determines the {\it lowest} possible energy density in this universe. As the energy density of normal matter drops below this value, the thermal ambience of the deSitter phase will remain constant and provide the irreducible `vacuum noise'. Note that the dark energy density is the the geometric mean $\rho_{DE}=\sqrt{\rho_\Lambda\rho_P}$ between the two energy densities. If we define a dark energy length scale $L_{DE}$  such that $\rho_{DE}=1/L_{DE}^4$ then $L_{DE}=\sqrt{L_PL_\Lambda}$ is the geometric mean of the two length scales in the universe. The figure \ref{fig:tpplumian} also shows the  $L_{DE}$ by broken horizontal lines. 
 
 While the two deSitter phases can last forever in principle, there is a natural cut off length scale in both of them
 which makes the region of physical relevance to be finite.\cite{plumian} In the the case of re-inflation in the late universe, this
 happens (at point F) when  the temperature of the CMBR radiation drops below the deSitter temperature.
 The universe will be essentially dominated by the vacuum thermal noise of the deSitter phase
 for $a>a_F$.
 One can easily determine the dynamic range of
 DF to be
 \begin{equation}
\frac{a_F}{a_D} \approx 2\pi T_0 L_\Lambda \left( \frac{\Omega_\Lambda}{\Omega_{DM}}\right)^{1/3}
\approx 3\times 10^{30}
\end{equation} 
 A natural bound on the  duration of inflation arises for a different reason. Consider a perturbation at some given wavelength scale which is stretched with the expansion of the universe as $\lambda\propto a(t)$.
 (See the line marked AB in Fig.\ref{fig:tpplumian}.) If there was no re-inflation, {\it all} the perturbations will `re-enter' the Hubble radius at some time (the point B in Fig.\ref{fig:tpplumian}).
 But if the universe undergoes re-inflation, then the Hubble radius `flattens out' at late times and some of the perturbations will {\it never} reenter the Hubble radius ! This criterion selects the portion of the inflationary phase (marked by CE) which
 can be easily calculated to be:
  \begin{equation}
\frac{a_{E} }{a_C} = \left( \frac{T_0 L_\Lambda}{T_{\rm reheat} H_{in}^{-1}}\right)
\left( \frac{\Omega_\Lambda}{\Omega_{DM}}\right)^{1/3}=\frac{(a_F/a_D)}{2\pi T_{\rm reheat} H_{in}^{-1}} \cong 10^{25}
\end{equation} 
where we have assumed a GUTs scale inflation with $E_{\rm GUT}=10^{14}$ GeV $\cong T_{\rm reheat}$ and $\rho_{in}=E_{\rm GUT}^4$
giving $2\pi H^{-1}_{in}T_{\rm reheat}=(3\pi/2)^{1/2}(E_P/E_{\rm GUT})\approx 10^5$.
For a Planck scale inflation with $2\pi H_{in}^{-1} T_{\rm reheat} = \mathcal{O} (1)$, the phases CE and DF are approximately equal. The region in the quadrilateral CEDF is the most relevant part of standard cosmology, though the evolution of the universe can extend to arbitrarily large stretches in both directions in time. This figure is definitely telling us something regarding the time translation invariance of the universe (`the perfect cosmological principle') and --- more importantly ---
\textit{about the  breaking of this symmetry}, and it deserves more attention than it has received.

Let us now turn  to several other features related to the \cc.
A \textit{non-representative} sample of attempts to understand/explain the \cc\ include
those based on QFT in curved space time,\cite{cc1}
those based on renormalisation group arguments,\cite{cc2}
quantum cosmological considerations,\cite{cc3}
 various cancellation mechanisms\cite{cc4} and many others. A study of these (failures!)
 reveals the following:

(a) If observed dark energy is due to \cc, we need to explain the small value of the dimensionless number $\Lambda(G\hbar/c^3)\approx 10^{-120}$.
The presence of $G\hbar$ clearly indicates that we are dealing with quantum mechanical problem, coupled to gravity. Any purely classical solution (like a classically decaying \cc)  will require (hidden or explicit) fine-tuning. At the same time, this is clearly an infra-red issue, in the sense that the phenomenon occurs at extremely low energies!.

(b) In addition to the zero-point energy of vacuum fluctuations (which \textit{must} gravitate\cite{tpijmpdrev}) the
phase transitions in the early universe (at least the well established electro-weak transition) change the ground state energy by a large factor. \textit{It is
necessary to arrange matters so that gravity does not respond to such changes.} Any approach to \cc\ which does not take this factor into account 
 is fundamentally flawed.

(c) An immediate consequence is that, the gravitational degrees of freedom which couple to \cc\ must have a special status and behave in a manner
different from other degrees of freedom. (The non linear coupling of matter with gravity has several subtleties; see eg. Ref.~\refcite{gravitonmyth}.) If, for example, we have a theory in which the source of gravity is
$(\rho +p)$ rather than $(\rho +3p)$, then \cc\ will not couple to gravity at all.  Unfortunately
it is not possible to develop a covariant theory of gravity using $(\rho +p)$ as the source. But we can achieve the same objective in different manner. Any metric $g_{ab}$ can be expressed in the form $g_{ab}=f^2(x)q_{ab}$ such that
${\rm det}\, q=1$ so that ${\rm det}\, g=f^4$. From the action functional for gravity
\begin{equation}
A=\frac{1}{2\kappa}\int \sqrt{-g}\,d^4x (R -2\Lambda)
=\frac{1}{2\kappa}\int \sqrt{-g}\,d^4x R -\frac{\Lambda}{\kappa}\int d^4x f^4(x)
\end{equation}
it is obvious that the \cc\ couples {\it only} to the conformal factor $f$. So if we consider a theory of gravity in which $f^4=\sqrt{-g}$ is kept constant and only $q_{ab}$ is varied, then such a model will be oblivious of
direct coupling to \cc\ and will not respond to changes in bulk vacuum energy. If the action (without the $\Lambda$ term) is varied, keeping ${\rm det}\, g=-1$, say, then one is lead to a {\it unimodular theory of gravity} with the equations of motion 
$R_{ab}-(1/4)g_{ab}R=\kappa(T_{ab}-(1/4)g_{ab}T)$ with zero trace on both sides. Using the Bianchi identity, it is now easy to show that this is equivalent to a theory with an {\it  arbitrary} \cc. That is, \cc\ arises as an (undetermined) integration constant in this model.\cite{unimod}
Unfortunately, we still need an extra physical principle to fix its value.

(d) The conventional discussion of the relation between cosmological constant and  the zero point energy is too simplistic since the zero point energy has no observable consequence. The observed non trivial features of the vacuum state arise from the {\it fluctuations} (or modifications) of this vacuum energy.
(This was, in fact,  known fairly early in the history of cosmological constant problem; see, e.g., Ref.\refcite{zeldo}).
If the vacuum probed by the gravity can readjust to take away the bulk energy density $\rho_P\simeq L_P^{-4}$, quantum \textit{fluctuations} can generate
the observed value $\rho_{\rm DE}$. One of the simplest models\cite{tpcqglamda} which achieves this uses the fact that, in the semiclassical limit, the wave function describing the universe of proper four-volume ${\cal V}$ will vary as
$\Psi\propto \exp(-iA_0) \propto 
 \exp[ -i(\Lambda_{\rm eff}\mathcal V/ L_P^2)]$. If we treat 
  $(\Lambda/L_P^2,{\cal V})$ as conjugate variables then uncertainty principle suggests $\Delta\Lambda\approx L_P^2/\Delta{\cal V}$. If
the four volume is built out of Planck scale substructures, giving $ {\cal V}=NL_P^4$, then the Poisson fluctuations will lead to $\Delta{\cal V}\approx \sqrt{\cal V} L_P^2$ giving
    $ \Delta\Lambda=L_P^2/ \Delta{\mathcal V}\approx1/\sqrt{{\mathcal V}}\approx   H_0^2
 $. (This idea can be made more quantitative\cite{tpcqglamda,volovikilya}.)

In fact, it is \textit{inevitable} that in a universe with two length scale $L_\Lambda,L_P$, the vacuum
 fluctuations will contribute an energy density of the correct order of magnitude $\rho_{DE}=\sqrt{\rho_\Lambda\rho_P}$. The hierarchy of energy scales in such a universe has\cite{plumian,tpvacfluc}
 the pattern
 \begin{equation}
\rho_{\rm vac}={\frac{1}{ L^4_P}}    
+{\frac{1}{L_P^4}\left(\frac{L_P}{L_\Lambda}\right)^2}  
+{\frac{1}{L_P^4}\left(\frac{L_P}{L_\Lambda}\right)^4}  
+  \cdots 
\label{energyhier}
\end{equation}  
 The first term is the bulk energy density which needs to be renormalized away (by a process which we  do not understand at present); the third term is just the thermal energy density of the deSitter vacuum state; what is interesting is that quantum fluctuations in the matter fields \textit{inevitably generate} the second term. A rigorous calculation\cite{tpvacfluc} of the dispersion in the energy shows that
the \textit{fluctuations} in the energy density $\Delta\rho$, inside
a region bounded by a cosmological horizon, is
 given by 
 \begin{equation}
 \Delta \rho_{\rm vac}   \propto L_P^{-2}L_\Lambda^{-2} \propto \frac{H_\Lambda^2}{G}
 \label{final}
 \end{equation}
 The numerical coefficient will depend on  the precise nature of infrared cutoff 
 radius (like whether it is $L_\Lambda$ or $L_\Lambda/2\pi$ etc.). 
 But {\it one cannot get away from} a fluctuation of magnitude $\Delta\rho_{vac}\simeq H_\Lambda^2/G$ that will exist in the
energy density inside a sphere of radius $H_\Lambda^{-1}$ if Planck length is the UV cut off. 
Since  observations suggest that there is indeed a $\rho_{vac}$ of \textit{similar} magnitude in the universe, it seems 
natural to identify the two, after subtracting out the mean value for reasons which we do not understand. This approach explains why there is a \textit{surviving} cosmological constant which satisfies 
$\rho_{DE}=\sqrt{\rho_\Lambda\rho_P}$ but not why the leading term in Eq.~(\ref{energyhier}) should be removed.

\section{Deeper Issues in Cosmology}

It is clear from the above discussion that `parametrised cosmology', which attempts to describe the evolution of the universe in terms of a small number of parameters, has made considerable progress in recent years. Having done this, it is  tempting to  ask 
more ambitious questions, some of which we will briefly discuss in this section.

There are two obvious questions a cosmologist faces every time (s)he gives a popular talk, for which (s)he has no answer!
The first one is: \textit{Why do the parameters of the universe have the values they have?} 
Today, we have no clue why  the real universe follows one template
out of a class of models all of which are permitted by the 
known laws of physics
 (just as we have no idea why there are three
families of leptons with specified mass ratios etc.)
Of the different cosmological parameters, $\Omega_{DM},\Omega_B,\Omega_R$ as well as the parameters of the initial power spectrum
$A,n$ should arise from  viable particle physics  models which actually says something about phenomenology. (Unfortunately, these research areas are not currently very fashionable.) On the other hand, it is not clear how we can understand $\Omega_{DE}$ 
without a reasonably detailed model for quantum gravity. \textit{In fact, the acid test for any viable quantum gravity model is whether it has something nontrivial to say about} $\Omega_{DE}$; all the current candidates have nothing to offer on this issue and thus fail the test.

The second question is: \textit{How (and why!) was the universe created and what happened before the big bang ?} The cosmologist giving the public lecture usually mumbles something about requiring a quantum gravity model to circumvent the classical singularity --- but we really have no idea!.
String theory offers no insight; the implications of loop quantum gravity for quantum cosmology 
have attracted fair mount of attention recently\cite{lqgqc} but it is 
 fair to say we still do not know how (and why) the universe came into being. 
 
 What is not often realised is that
 \textit{certain aspects of this problem transcends the question of technical tractability of quantum gravity} and can be presented in more general terms. Suppose, for example, one has produced some kind of theory for quantum gravity. Such a theory is likely to come with a single length scale $L_P$.
 Even when one has the back drop of such a theory, it is not clear how  one hopes to address questions like:
 (a) Why is our universe much bigger than $L_P$ which is the only scale in the problem i.e., why is the mean curvature of the universe
 much smaller than $L_P^{-2}$?  (b) How   does
the universe, treated as a dynamical system,  evolve \textit{spontaneously} from a quantum regime to classical regime? (c) How does one obtain the notion of a cosmological arrow of time, starting from timeless or at least time symmetric description?  

One viable idea regarding these issues
seems to be  based on vacuum instability which describes the universe as an unstable system with  an unbounded Hamiltonian. Then it is  possible for
the expectation value of spatial curvature to vary monotonically as, say, $<R>\propto L_P^{-2}(t/t_P)^{-\alpha}$ with some index $\alpha$, as the universe expands in an unstable mode. Since the conformal factor of the metric has the `wrong' sign for the kinetic energy term, this mode will become semiclassical first. Even then it is not clear how the arrow of time related to the expanding phase of the universe arises; one needs to invoke decoherence like arguments to explain the classical limit\cite{semicosmo} and the situation is not very satisfactory.

While the understanding of such `deeper' issues might require the \textit{details} of the viable model for quantum gravity, one should not ignore the  alternative possibility that we are just not clever enough. It may turn out that  certain obvious (low energy) features of the universe, that we take for granted, contain clues to  the full theory of quantum gravity
 (just as the equality of inertial and gravitational masses, known for centuries, was
 turned on its head by Einstein's insight)
 if only we manage to find the right questions to ask. To illustrate this point, consider an atomic physicist who solves the Schrodinger equation for the electrons in the helium atom. (S)he will discover that, of all the permissible solutions, only half
(which are antisymmetric under the exchange of electrons) are realized in nature though the Hamiltonian of the helium atom offers no insight for this feature. This is a low energy phenomenon
the explanation of which lies deep inside relativistic field theory. 

In this spirit, there are  at least two peculiar features of our universe  which are noteworthy: 

(i) The first one, which is  already mentioned,
is the fact that our universe seemed to have evolved \textit{spontaneously} from a quantum regime to classical regime bringing with it the notion of a cosmological arrow of time.\cite{semicosmo} This
is not a generic feature of dynamical systems (and is connected with the fact that the Hamiltonian for the system is unbounded from below). 

(ii)
The second issue corresponds to the low energy vacuum state of matter fields in the universe and the notion of the particle ---
which is an excitation around this vacuum state. Observations show that, in the classical limit, we do have the notion of an inertial frame,  vacuum state and a notion of the particle such that the particle at rest in this frame will see, say, the CMBR as isotropic. This is nontrivial, since the notion of classical particle arises after several limits are taken: Given the formal quantum state
$\Psi[g,matter]$ of gravity and matter, one would first proceed to a limit of quantum fields in curved background, in which
the gravity is treated as a c-number\cite{semigrav}. Further taking the $c\to\infty$ limit (to obtain quantum mechanics) and $\hbar\to0$ limit (to obtain the classical limit) one will reach the notion of a particle in classical theory. Miraculously enough, the full quantum state of the universe seems to finally  select (in the low energy classical limit) 
a local inertial frame, such that the particle at rest in this frame will see the universe as isotropic --- rather than the universe as accelerating or rotating, say. 
This is a nontrivial constraint on $\Psi[g,matter]$, especially since the vacuum state and particle concept are ill defined in the curved background\cite{probe}. One can
show that this feature imposes special conditions on the wave function of the universe\cite{choices} in simple minisuperspace models but
its wider implications in quantum gravity are unexplored.

\end{document}